\documentclass[a4paper]{article}
\usepackage[ansinew]{inputenc}
\usepackage{times}
\usepackage[T1]{fontenc}
\usepackage{graphicx}
\usepackage{geometry}
\usepackage{amssymb}
\usepackage{amsmath}

\usepackage{rotating}

\usepackage{xcolor}
\colorlet{shadecolor}{gray!25}
\usepackage{subfig}
\usepackage{float}

\begin{document}
\title{A Tukey type trend test for repeated carcinogenicity bioassays, motivated by multiple glyphosate studies}
\author{Ludwig A. Hothorn and Frank Schaarschmidt, \\
Natural Sci. Faculty, Leibniz University, Herrenhauserstr. 2, 30419 Hannover, Germany}
\maketitle

\begin{abstract}
In the last two decades, significant methodological progress to the simultaneous inference of simple and complex randomized designs, particularly proportions as endpoints, occurred. This includes: i) the new Tukey trend test approach \cite{Schaarschmidt2019}, ii) multiple contrast tests for binomial proportions \cite{Schaarschmidt2008}, iii) multiple contrast tests for poly-k estimates \cite{Schaarschmidt2008a}, and Add-1 approximation for one-sided inference \cite{Schaarschmidt2014}. This report focus on a new Tukey type trend test to evaluate repeated long-term carcinogenicity bioassays which was motivated by multiple glyphosate studies. Notice, it is not the aim here to contribute to the evaluation of Glyphosate and its controversies. By means of the CRAN-packages \verb|tukeytrend, MCPAN, multcomp| the real data analysis is straightforward possible.
\end{abstract}

\section{Introduction}
Trend tests on tumor-site specific crude proportions $ n_{\text{developing a tumor}}/ n_{\text{at risk}}$ and their mortality-adjusted poly-k estimates are recommended for the analysis of a potential tumorigenic effect of a compound in long-term carcinogenicity bioassays on rodents \cite{NTP2013}. Repeated studies are common: i) for rats and mice, ii) for males and females, iii) for relevant chemicals by several sponsors, e.g. glyphosate bioassays with 5 mice and 9 rats studies \cite{Greim2015}, iv) in earlier days to compensate extensive mortality by an improved second design. \\
Up to now, an independent evaluation of each individual study has been common practice. From a regulatory risk assessment perspective, a joint evaluation may be appropriate, particularly it is only one single decision instead of having more than one. There does not seem to be a viable approach to this yet. The approaches currently used, a simple summary  into a single pooled contingency table \cite{Greim2015} or the categorization into low, median, high dose over in fact  wide dose ranges \cite{Portier2016} such as for the Glyphosate studies, seems to be inappropriate. The glyphosate example below illustrates the statistical problems where the crude proportions for 5 studies (abbreviated by A-D) for malign lymphoma in male mice are considered \cite{Tarazona2017}, see Fig. 1. 

\begin{figure}[ht]
	\centering
		\includegraphics[width=0.65\textwidth]{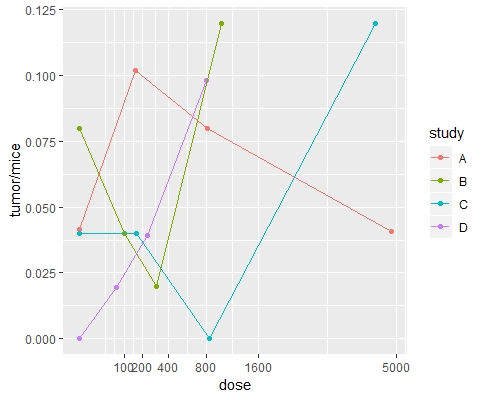}
	\label{fig:tarazona}
	\caption{Dose-dependent crude proportions of malign lymphoma in male mice in 5 studies with glyphosate- an example}
	\end{figure}

 First, the dose ranges of individual studies vary by orders of magnitude. If the Paracelsus theorem is to be taken seriously, the dose must be modeled precisely, i.e. quantitatively. Second, the shapes of dose-response relationships are very different between studies and between tumor types. Therefore, a trend test for near-to-linear trends only, such as the recommended Armitage test \cite{Armitage1955} (and its poly-3 modification), does not seem to be particularly suitable.  Thirdly, mortality patterns between dose groups and studies may be significantly different so that mortality-adjusted analysis, such as simple  poly-k-test \cite{Bailer1988}, may also be available.\\
Therefore, a Tukey type trend test for crude proportion and poly-k estimates \cite{Hothorn2019} within a mixed effect model with the random factor \textit{between studies} is proposed.

\section{A Tukey type trend test on crude proportions for repeated studies}
The Tukey trend test \cite{Tukey1985} can be formulated as a maximum test $T^{max}=max(T_1,T_2,T_3)$, with the 3 regression tests for linear, ordinal and logarithmic dose scores. The global $T^{max}$ test is jointly k-variate normal-distributed where the common correlation matrix is hard to calculate. The multiple marginal models approach (mmm) \cite{Pipper2012} provides the joint distribution of parameter estimates where the variance-covariance matrix of parameter estimates is obtained using derivatives of the log likelihood function, i.e. without the explicit formulation of the correlation matrix in linear, generalized linear and generalized linear mixed effects models \cite{Schaarschmidt2019}. Its modification for proportions base on the estimates of generalized linear models (glm) where the canonical link is the logit function with the odds ratio as effect size. This asymptotic approaches reveal problems to keep the nominal coverage probability for small sample sizes, and the common used-NTP-design with $n_i=50,50,50,50$ is already \textit{small} under this perspective. Furthermore, ratio as effect size for multiple contrasts is rather unstable when considering $p_0=0$ which is a  relevant data outcome in long-term carcinogenicity bioassays. Both phenomena can be mitigated to some extend by Add-2 data transformation \cite{Agresti2000}, simply adding one pseudo count to both tumor and no-tumor category in each dose group. For one-sided inference, and in carcinogenicity bioassays only increasing tumor rates are of interest, an Add-1 approximation \cite{Schaarschmidt2014} behaves even better. \\
Several approaches can be used to account for the between-studies variability: i) mixed effect model, ii) combining the p-values of independent studies by e.g. Fishers product criterion, iii) fixed effect two-way layout model and iv) naive pooled contingency table approach. Here only the first approach, seemingly the most appropriate, is discussed in detail without the aim of method comparisons.\\
Penalized quasi-likelihood generalized linear mixed models (glmm) allowing the modeling of study-specific intercepts $p_{C, i}$ and slopes $b_i$, e.g. by the function \verb|glmmPQL| in library(MASS) \cite{MASS}. This approach yields downwardly biased variance estimates for binomial glmm, but for not too small sample sizes this bias is tolerable \cite{Rich2018}. The following three models are used with the Add-1 approximation to achieve one-sided lower inference for small $n_i$ and possible $p_0=0$, each for an arithmetic, ordinal or logarithmic dose score:

\scriptsize
\begin{verbatim}
library(tukeytrend)
Nmic <- dosescalett(Nmice, dose="dose", scaling=c("ari", "ord", "arilog"))$data
glmmari <- glmmPQL(fixed=cbind(tumor+0.5,(rats-tumor)+0.5) ~ doseari, random = ~ 1 +doseari |Study,
                   family = binomial, data=Nmic)
glmmord <- glmmPQL(fixed=cbind(tumor+0.5,(rats-tumor)+0.5) ~ doseord, random = ~ 1 +doseari |Study,
                   family = binomial, data=Nmic, niter = 100)
glmmarilog <- glmmPQL(cbind(tumor+0.5,(rats-tumor)+0.5) ~ dosearilog, random = ~ 1 +doseari |Study,
                   family = binomial, data=Nmic)
\end{verbatim}
\normalsize
The function \verb|lmer2lm| allows the transformation of the mixed effect estimates into fixed effect estimates \cite{Ritz2017}. Again, the functions \verb|mmm, mlf| provide the 3-variate normal distribution for the estimated variance-covariance matrix of these 3 pseudo-fixed-effect models: 

\scriptsize
\begin{verbatim}
lmari <- tukeytrend:::lmer2lm(glmmari)
lmord <- tukeytrend:::lmer2lm(glmmord)
lmarilog <- tukeytrend:::lmer2lm(glmmarilog)
linf <- matrix(c(0,1), ncol=2)
ttglmm <- glht(mmm("mari"=lmari, "mord"=lmord, "marilog"=lmarilog),
               mlf("mari"=linf, "mord"=linf, "marilog"=linf), alternative="greater")
summary(ttglmm)
\end{verbatim}
\normalsize
None of the three models are in the alternative in this joint analysis whereas by the independent analysis the studies no. C and D ($p_C^{ari}=0.027$ and $p_D^{ari}=0.014$) are. This is not a surprising effect. Although  in the joint analysis the p-value decreases with increasing sample sizes, the p-value increases with increased variance components for either heterogeneity of studies and/or slope as well as with  different shapes in the different regression models. And in the above example, the shapes are rather different.

\section{A Tukey type trend test on poly-k estimates for repeated studies}
Dose-specific mortality $t_{ij}$ (as time of death of animal $j$ in dose $i$) can be accounted by individual weights $w_{ij}=\left( t_{ij}/t_{max} \right)^k$ reflecting individual mortality pattern within the poly-k adjustment \cite{Bailer1988}. Now, the penalized quasi-likelihood glm is extended to a weighted formulation in the function \verb|glmmPQL|.\\
\newline
We were unable to find repeated studies on a substance with both individual tumor and mortality data. Therefore, the males and females rat study for zymbal tumors (adenoma or carcinoma) in the NTP TR-365 on  \cite{NTP365} were used as toy examples, where sex is considered as random factor accordingly.
\footnotesize
\begin{verbatim}
library(tukeytrend)
library(MASS)
TN1 <- dosescalett(ZYM, dose="dose", scaling=c("ari", "ord", "arilog"))$data
glmmari1T <- glmmPQL(fixed=zymbal ~ doseari, random = ~ 1 |sex,weight=weightpoly3,
                   family = binomial, data=TN1)
glmmord1T <- glmmPQL(fixed=zymbal ~ doseord, random = ~ 1 |sex,weight=weightpoly3,
                   family = binomial, data=TN1, niter = 100)
glmmarilog1T <- glmmPQL(fixed=zymbal ~ dosearilog, random = ~ 1 |sex,weight=weightpoly3,
                   family = binomial, data=TN1)
lmari1T <- tukeytrend:::lmer2lm(glmmari1T)
lmord1T <- tukeytrend:::lmer2lm(glmmord1T)
lmarilog1T <- tukeytrend:::lmer2lm(glmmarilog1T)
linf <- matrix(c(0,1), ncol=2)
mm1T <- glht(mmm("mari"=lmari1T, "mord"=lmord1T, "marilog"=lmarilog1T),
               mlf("mari"=linf, "mord"=linf, "marilog"=linf), alternative="greater")
Mt1T<-summary(mm1T)
\end{verbatim}
\normalsize
Here, the similarity of both dose-response relationship yield in a significant p-value for the joint trend test ($p=0.0125)$ whereas the consideration males and females studies alone, none is significant ($p^F=0.060, p^M=0.127$). The shapes for males and females do not differ to much.

\section{Conclusion}
With the new trend test, either crude tumor rates or their mortality-adjusted poly-k estimates can be used for the joint evaluation of repeated carcinogenicity bioassays. It combines the Tukey-test principle with the poly-k and Add-1 modifications. Related R-code is available using the packages \verb|tukeytrend|, \verb|multcomp,MASS, MCPAN|. Both adjusted p-values and simultaneous confidence intervals are available for the three common effect sizes risk difference, risk ratio and odds ratio (only the latter explicitly shown). This approach can be extended to the consideration of multiple tumors jointly and the optimal choice of the tuning parameter $k$ in the poly-k adjustment. \\ More work is needed, particularly simulation studies for method comparisons to characterize size and power behavior, particularly for unbalanced small sample size designs.

\footnotesize
\bibliographystyle{plain}

\section{Appendix: R-code}
\tiny
\begin{verbatim}
<<multipleS1, echo=FALSE, results='asis', fig.keep='none', message=FALSE, warning=FALSE>>=
###################### glyphosat tumors
library(ggplot2)
library(tukeytrend)
library(lme4)
library("MASS")
library(xtable)

Nmice <- data.frame(
  study=c(rep("Ex1",8), rep("Ex2",8)),
  sex=c("m", "m", "m", "m", "f", "f", "f", "f"),
  dose = c(0, 98, 299, 1081, 0,15,151,1460,
           0, 153, 787, 2116, 0,100,300, 1000),
  tumor = c(8,8,10,11, 11,20,19,19,4,4,8,7,8,12,9,10),
  rats = c(50,50,50,45,40,40,35,30, 60,60,60,50,25,25,20,20))

NMI<-xtable(Nmice, digits=c(0,1,1,0,0,0))
print(NMI, include.rownames=FALSE, 
      caption.placement = "top",size="\\footnotesize")
@

\item This toy example is strange because of study-specific: i) control rates, ii) shapes of dose-response, iii) sample sizes, and heterogeneity between all 4 studies, iv) plateau effect

<<multipleS2, echo=FALSE, result="hide", out.width='.39\\linewidth', message=FALSE, warning=FALSE, fig.pos="H">>=
Nmice$Study<-Nmice$study:Nmice$sex
ggplot(Nmice, aes(y=tumor/rats, x=dose))  + geom_point(aes(color=Study)) +
  geom_line(aes(color=Study, group=Study)) +
  scale_x_continuous(trans="sqrt",breaks=c(0,100,200,400,800,1600,5000))
@

\item Approach I: mixed effect model using add1 approximation, allowing study-specific intercepts= $p_{C, i}$ and slopes $b_i$. Standard logistic regression with odds ratio as effect size. One-sided fro an increase. Global best model with estimated slope (and its CI or p-value). Tendency: $\Downarrow$ p-value with $\Uparrow$ $n_i$ and/or $\Downarrow$ variance components for either heterogeneity of  $p_{C, i}$ and/or slopes $b_i$ and complex influence of different shapes (= different regression models).
\tiny

<<multipleS3add, echo=TRUE, result="hide", message=FALSE, warning=FALSE>>=
library(tukeytrend)
Nmic <- dosescalett(Nmice, dose="dose", scaling=c("ari", "ord", "arilog"))$data

glmmari <- glmmPQL(fixed=cbind(tumor+0.5,(rats-tumor)+0.5) ~ doseari, random = ~ 1 +doseari |Study,
                   family = binomial, data=Nmic)
glmmord <- glmmPQL(fixed=cbind(tumor+0.5,(rats-tumor)+0.5) ~ doseord, random = ~ 1 +doseari |Study,
                   family = binomial, data=Nmic, niter = 100)
glmmarilog <- glmmPQL(cbind(tumor+0.5,(rats-tumor)+0.5) ~ dosearilog, random = ~ 1 +doseari |Study,
                   family = binomial, data=Nmic)
lmari <- tukeytrend:::lmer2lm(glmmari)
lmord <- tukeytrend:::lmer2lm(glmmord)
lmarilog <- tukeytrend:::lmer2lm(glmmarilog)
linf <- matrix(c(0,1), ncol=2)
ttglmm <- glht(mmm("mari"=lmari, "mord"=lmord, "marilog"=lmarilog),
               mlf("mari"=linf, "mord"=linf, "marilog"=linf), alternative="greater")
a2<-summary(ttglmm)
@
\normalsize

\item Approach II: independent individual studies

<<multipleS4, echo=TRUE, results='asis', message=FALSE, warning=FALSE>>=
# per study separate add1
NmicA<- droplevels(subset(Nmice, study =="Ex1"))
NmicAm<- droplevels(subset(NmicA, sex =="m"))
NmicAf<- droplevels(subset(NmicA, sex =="f"))

ATf <- glm(cbind(tumor+0.5,rats-tumor+0.5) ~ dose,  data=NmicAf, family=binomial())
caf <- tukeytrendfit(ATf, dose="dose", scaling=c("ari", "ord", "arilog"))
ex1f<-summary(asglht(caf,alternative="greater")) 

ATm <- glm(cbind(tumor+0.5,rats-tumor+0.5) ~ dose,  data=NmicAm, family=binomial())
cam <- tukeytrendfit(ATm, dose="dose", scaling=c("ari", "ord", "arilog"))
ex1m<-summary(asglht(cam,alternative="greater"))   

NmicW<- droplevels(subset(Nmice, study =="Ex2"))
NmicWm<- droplevels(subset(NmicW, sex =="m"))
NmicWf<- droplevels(subset(NmicW, sex =="f"))

WTm <- glm(cbind(tumor+0.5,rats-tumor+0.5) ~ dose,  data=NmicWm, family=binomial())
cwm <- tukeytrendfit(WTm, dose="dose", scaling=c("ari", "ord", "arilog"))
ex2m<-summary(asglht(cwm,alternative="greater"))      

WTf <- glm(cbind(tumor+0.5,rats-tumor+0.5) ~ dose,  data=NmicWf, family=binomial())
cwf <- tukeytrendfit(WTf, dose="dose", scaling=c("ari", "ord", "arilog"))
ex2f<-summary(asglht(cwf,alternative="greater"))      
@

\item Approach III: Fixed effect model

<<fixed1, echo=TRUE, results='asis', message=FALSE, warning=FALSE>>=
fitc <- glm(cbind(tumor+0.5,(rats-tumor)+0.5) ~ dose+Study,family = binomial, data=Nmic)
covars<-anova(fitc)
ttc<- tukeytrendfit(fitc, dose="dose", scaling=c("ari", "ord", "arilog"))
fix<-summary(asglht(ttc, alternative="greater"))
@

\item Approach IV: Naive pooled data

<<fixePool, echo=TRUE, results='asis', message=FALSE, warning=FALSE>>=
fitp <- glm(cbind(tumor+0.5,(rats-tumor)+0.5) ~ dose,family = binomial, data=Nmic)
covars<-anova(fitp)
ttp<- tukeytrendfit(fitp, dose="dose", scaling=c("ari", "ord", "arilog"))
pool<-summary(asglht(ttp, alternative="greater"))
@


\item Approach V: Fisher product criterion for log-lin trends

<<fish, echo=TRUE, results='asis', warning=FALSE, message=FALSE, warning=FALSE>>=
library(metap)
fisp<-cbind(ex1f$test$pvalue[3], ex1m$test$pvalue[3], ex2f$test$pvalue[3], ex2m$test$pvalue[3])
fis<-sumlog(fisp)
@

\item Conclusions: mixed effect model more appropriate than all other possible approaches

<<suma, echo=FALSE, results='asis', message=FALSE, warning=FALSE>>=
library("ggplot2")
Ex1m <-fortify(ex1m)[, 6]
Ex2m <-fortify(ex2m)[, 6]

Ex1f <-fortify(ex1f)[, 6]
Ex2f <-fortify(ex2f)[, 6]


Po <-fortify(pool)[, 6]
Fi <-fortify(fix)[, 6]
Mi <-fortify(a2)[, c(1,6)]
all<-cbind(Mi, Ex1f,Ex1m,Ex2f, Ex2m,Po,Fi,fis$p )
colnames(all) <-c("Model", "Mixed", "Only1m", "Only1f", "Only2m", "Only2f", "Pooled", "Fixed", "Fisher")
print(xtable(all, digits=4, caption="Methdod comparison", 
             label="tab:IArl"), include.rowname=FALSE, caption.placement = "top", sanitize.text.function = function(x){x}, size="footnotesize")

@
<<mic1, echo=FALSE, results='asis', fig.keep='none', message=FALSE, warning=FALSE>>=
###################### glyphosat tumors
library(ggplot2)
library(tukeytrend)
library(lme4)
library("MASS")
library(xtable)

Z365 <- data.frame(
  study=c(rep("A",3), rep("B",3)),
  dose = c(0, 25, 50, 0,25,50),
          Wscore=c("C","D1","D2","C","D1","D2"),
  tumor = c(0,3,2,0,1,3),
  mice = c(49,45,41,36,37,35))

Lmice<-Z365

Lmice <- data.frame(
  study=c(rep("A",4), rep("B",4),rep("C",4), rep("D",4)),
  dose = c(0, 157, 814, 4841, 0,100,300,1000,
           0, 165, 838, 4348, 0,71,234, 810),
  Wscore=c("C","D1","D2","D2", "C","D1","D2","D2", "C","D1","D2","D2", "C","D1","D2","D3"),
  tumor = c(2,5,4,2, 4,2,1,6,2,2,0,6,0,1,2,5),
  mice = c(48,49,50,49,50,50,50,50, 50,50,50,50,51,51,51,51))

LMI<-xtable(Lmice, digits=c(0,1,0,0,0,0))
print(LMI, include.rownames=FALSE, 
      caption.placement = "top",size="\\footnotesize")

@

<<mice2, echo=FALSE, result="hide", out.width='.39\\linewidth', message=FALSE, warning=FALSE, fig.pos="H">>=
ggplot(Lmice, aes(y=tumor/mice, x=dose))  + geom_point(aes(color=study)) +
  geom_line(aes(color=study, group=study)) +
  scale_x_continuous(trans="sqrt",breaks=c(0,100,200,400,800,1600,5000))
@



<<mice3, echo=TRUE, result="hide", message=FALSE, warning=FALSE>>=
library(tukeytrend)
Lmic <- dosescalett(Lmice, dose="dose", scaling=c("ari", "ord", "arilog"))$data

glmmari <- glmmPQL(fixed=cbind(tumor,mice-tumor) ~ doseari, random = ~ 1 +doseari |study,
                   family = binomial, data=Lmic)
glmmord <- glmmPQL(fixed=cbind(tumor,mice-tumor) ~ doseord, random = ~ 1 +doseari |study,
                   family = binomial, data=Lmic, niter = 100)
glmmarilog <- glmmPQL(cbind(tumor,mice-tumor) ~ dosearilog, random = ~ 1 +doseari |study,
                   family = binomial, data=Lmic)
lmari <- tukeytrend:::lmer2lm(glmmari)
lmord <- tukeytrend:::lmer2lm(glmmord)
lmarilog <- tukeytrend:::lmer2lm(glmmarilog)
linf <- matrix(c(0,1), ncol=2)
ttglmm <- glht(mmm("mari"=lmari, "mord"=lmord, "marilog"=lmarilog),
               mlf("mari"=linf, "mord"=linf, "marilog"=linf), alternative="greater")
l2<-summary(ttglmm)
@
\normalsize

\item Approach II: independent individual studies

<<mice4, echo=TRUE, results='asis', message=FALSE, warning=FALSE>>=
# per study separate add1
LmicA<- droplevels(subset(Lmice, study =="A"))
LmicB<- droplevels(subset(Lmice, study =="B"))
LmicC<- droplevels(subset(Lmice, study =="C"))
LmicD<- droplevels(subset(Lmice, study =="D"))

mA <-glm(cbind(tumor,mice-tumor)~dose, data=LmicA, family= binomial(link="logit"))
tA <- tukeytrendfit(mA, dose="dose", scaling=c("ari", "ord", "arilog"))
TA <- summary(glht(model=tA$mmm, linfct=tA$mlf, alternative="greater"))

mB <-glm(cbind(tumor,mice-tumor)~dose, data=LmicB, family= binomial(link="logit"))
tB <- tukeytrendfit(mB, dose="dose", scaling=c("ari", "ord", "arilog"))
TB <- summary(glht(model=tB$mmm, linfct=tB$mlf, alternative="greater"))

mC <-glm(cbind(tumor,mice-tumor)~dose, data=LmicC, family= binomial(link="logit"))
tC <- tukeytrendfit(mC, dose="dose", scaling=c("ari", "ord", "arilog"))
TC <- summary(glht(model=tC$mmm, linfct=tC$mlf, alternative="greater"))

mD <-glm(cbind(tumor,mice-tumor)~dose, data=LmicD, family= binomial(link="logit"))
tD <- tukeytrendfit(mD, dose="dose", scaling=c("ari", "ord", "arilog"))
TD <- summary(glht(model=tD$mmm, linfct=tD$mlf, alternative="greater"))

@

\item Approach III: Fixed effect model
Curious mean dose levels
<<mice5, echo=TRUE, results='asis', message=FALSE, warning=FALSE>>=
mF <-glm(cbind(tumor,mice-tumor)~dose+study, data=Lmic, family= binomial(link="logit"))
ttf<- tukeytrendfit(mF, dose="dose", scaling=c("ari", "ord", "arilog"))
lfix<-summary(asglht(ttf, alternative="greater"))
@


\item Approach IV: Williams teype trend by dose levels

<<mice5a, echo=TRUE, results='asis', message=FALSE, warning=FALSE>>=
mW <-glm(cbind(tumor,mice-tumor)~Wscore+study, data=Lmic, family= binomial(link="logit"))
Wfix<-summary(glht(mW, linfct = mcp(Wscore = "Williams"), alternative="greater"))
@


\item Approach IVa: Williams teype trend by dose levels, but pooled

<<mice5b, echo=TRUE, results='asis', message=FALSE, warning=FALSE>>=
mWp <-glm(cbind(tumor,mice-tumor)~Wscore, data=Lmic, family= binomial(link="logit"))
Wfp<-summary(glht(mWp, linfct = mcp(Wscore = "Williams"), alternative="greater"))
@


\item Approach V: Naive pooled data

<<mice6, echo=TRUE, results='asis', message=FALSE, warning=FALSE>>=
mP <-glm(cbind(tumor,mice-tumor)~dose, data=Lmic, family= binomial(link="logit"))
ttp<- tukeytrendfit(mP, dose="dose", scaling=c("ari", "ord", "arilog"))
lpool<-summary(asglht(ttp, alternative="greater"))
@


\item Approach VI: Fisher product criterion for log-lin trends

<<mice7, echo=TRUE, results='asis', warning=FALSE, message=FALSE, warning=FALSE>>=
library(metap)
fisp<-cbind(TA$test$pvalue[3], TB$test$pvalue[3], TC$test$pvalue[3], TD$test$pvalue[3])
lfis<-sumlog(fisp)
@

\item Conclusions: mixed effect model more appropriate than all other possible approaches

<<mice8, echo=FALSE, results='asis', message=FALSE, warning=FALSE>>=
library("ggplot2")
lA <-fortify(TA)[, 6]
lB <-fortify(TB)[, 6]
lC <-fortify(TC)[, 6]
lD <-fortify(TD)[, 6]
lPo <-fortify(lpool)[, 6]
lFi <-fortify(lfix)[, 6]
lMi <-fortify(l2)[, c(1,6)]
lW <-fortify(Wfix)[, c(1,6)]
lWp <-fortify(Wfp)[, c(1,6)]
lall<-cbind(lMi, lA,lB,lC,lD,lPo,lFi,lfis$p, lW[1,2], lWp[1,2] )
colnames(lall) <-c("Model", "Mix", "OnlyA", "OnlyB", "OnlyC", "OnlyD", "Pool", "Fix", "Fish", "Wil", "WiP")
print(xtable(lall, digits=3, caption="Method comparison mal. lymphoma", 
             label="tab:IArll"), include.rowname=FALSE, caption.placement = "top", sanitize.text.function = function(x){x}, size="footnotesize")

@

%%%%%%%%%%%%%%%%%%%%%%%%%%%%%%%%%%%%%%%%%%%%%%%%%%%%%%%%%%%%%%%%%%%%
\section{ poly3- Tukey}

Zymbal adenoma or carcinoma in TR365 for male and female rats


<<PolyT1, echo=FALSE, results='asis', message=FALSE, warning=FALSE>>=
setwd("D:/PUB/_MICRO2016TrendCarcino")
zym<-read.table("zymbaltr365.csv", sep=";", dec="," ,header=TRUE)
zymF<-droplevels(zym[zym$sex=="female", ])
zymM<-droplevels(zym[zym$sex=="male", ])
table(zymF$dose, zymF$zymbal)
table(zymM$dose, zymM$zymbal)
@

<<PolyT2, echo=TRUE, results='asis', message=FALSE, warning=FALSE>>=
library(MCPAN)
library(lme4)
library(multcomp)
library(tukeytrend)

# study-specific poly3 weights
zymF$weightpoly3 <- 1 # Compute the poly-3 (-k)- weights at the level of single animals
wt0f <- which(zymF$zymbal == 0)
zymF$weightpoly3[wt0f] <- (zymF$time[wt0f]/max(zymF$time))^3

zymM$weightpoly3 <- 1 # Compute the poly-3 (-k)- weights at the level of single animals
wt0m <- which(zymM$zymbal == 0)
zymM$weightpoly3[wt0m] <- (zymM$time[wt0m]/max(zymM$time))^3

ZYM<-rbind(zymM,zymF)
#ZYM$Dose<-as.factor(ZYM$dose)
# glmm with weights
#mz1 <- glmer(zymbal~ dose +(1|sex),family = binomial,
#             weight=weightpoly3, data=ZYM)
#Mz1 <- tukeytrendfit(mz1, dose="dose", scaling=c("ari", "ord", "arilog")) ### FRANK etwas falsch
#EXA17m<-summary(glht(model=Exa17m$mmm, linfct=Exa17m$mlf))

@

<<PolyT3, echo=TRUE, results='asis', message=FALSE, warning=FALSE>>=
library(tukeytrend)
library(MASS)
TN1 <- dosescalett(ZYM, dose="dose", scaling=c("ari", "ord", "arilog"))$data
glmmari1T <- glmmPQL(fixed=zymbal ~ doseari, random = ~ 1 |sex,weight=weightpoly3,
                   family = binomial, data=TN1)
glmmord1T <- glmmPQL(fixed=zymbal ~ doseord, random = ~ 1 |sex,weight=weightpoly3,
                   family = binomial, data=TN1, niter = 100)
glmmarilog1T <- glmmPQL(fixed=zymbal ~ dosearilog, random = ~ 1 |sex,weight=weightpoly3,
                   family = binomial, data=TN1)
lmari1T <- tukeytrend:::lmer2lm(glmmari1T)
lmord1T <- tukeytrend:::lmer2lm(glmmord1T)
lmarilog1T <- tukeytrend:::lmer2lm(glmmarilog1T)
linf <- matrix(c(0,1), ncol=2)
mm1T <- glht(mmm("mari"=lmari1T, "mord"=lmord1T, "marilog"=lmarilog1T),
               mlf("mari"=linf, "mord"=linf, "marilog"=linf), alternative="greater")
Mt1T<-summary(mm1T)
@

<<PolyT3f, echo=TRUE, results='asis', message=FALSE, warning=FALSE>>=
library(tukeytrend)
TN1f<- dosescalett(zymF, dose="dose", scaling=c("ari", "ord", "arilog"))$data
ari1Tf <- glm(zymbal ~ doseari, weight=weightpoly3,
                 family = binomial, data=TN1f)
ord1Tf <- glm(zymbal ~ doseord, weight=weightpoly3,
                   family = binomial, data=TN1f)
arilog1Tf <- glm(zymbal ~ dosearilog,weight=weightpoly3,
                   family = binomial, data=TN1f)

mm1Tf <- glht(mmm("mari"=ari1Tf, "mord"=ord1Tf, "marilog"=arilog1Tf),
               mlf("mari"=linf, "mord"=linf, "marilog"=linf), alternative="greater")
Mt1Tf<-summary(mm1Tf)
@

<<PolyT3m, echo=TRUE, results='asis', message=FALSE, warning=FALSE>>=
library(tukeytrend)
TN1m<- dosescalett(zymM, dose="dose", scaling=c("ari", "ord", "arilog"))$data
ari1Tm <- glm(zymbal ~ doseari, weight=weightpoly3,
                 family = binomial, data=TN1m)
ord1Tm <- glm(zymbal ~ doseord, weight=weightpoly3,
                   family = binomial, data=TN1m)
arilog1Tm <- glm(zymbal ~ dosearilog,weight=weightpoly3,
                   family = binomial, data=TN1m)

mm1Tm <- glht(mmm("mari"=ari1Tm, "mord"=ord1Tm, "marilog"=arilog1Tm),
               mlf("mari"=linf, "mord"=linf, "marilog"=linf), alternative="greater")
Mt1Tm<-summary(mm1Tm)
@


%%%%%%%%%%%%%%%%%%%%%%%%%%%%%%%%%%%%%%%%%%%%%%%%%%%%%%
\section{A poly3- Dunnett-type test for repeated studies}

\begin{itemize}
\item For all carcinogenicity studies with competing tumor progression and mortality, poly3 adjusted analysis is needed \cite{Kodell2012}
\item For repeated studies such an approach seems to be not available yet
\item  A version:
\tiny
<<fakedat, echo=FALSE, results='asis', message=FALSE, warning=FALSE>>=
#options(max.print=100000)
#setwd("E:/externals/Dow2017")
#meLH<-read.table("meLH.csv", sep=",", header=TRUE)# faked methly example
#dump("meLH", file="meLH.R")
meLH <-
structure(list(no = c(1L, 2L, 3L, 4L, 5L, 6L, 7L, 8L, 9L, 10L, 
11L, 12L, 13L, 14L, 15L, 16L, 17L, 18L, 19L, 20L, 21L, 22L, 23L, 
24L, 25L, 26L, 27L, 28L, 29L, 30L, 31L, 32L, 33L, 34L, 35L, 36L, 
37L, 38L, 39L, 40L, 41L, 42L, 43L, 44L, 45L, 46L, 47L, 48L, 49L, 
50L, 51L, 52L, 53L, 54L, 55L, 56L, 57L, 58L, 59L, 60L, 61L, 62L, 
63L, 64L, 65L, 66L, 67L, 68L, 69L, 70L, 71L, 72L, 73L, 74L, 75L, 
76L, 77L, 78L, 79L, 80L, 81L, 82L, 83L, 84L, 85L, 86L, 87L, 88L, 
89L, 90L, 91L, 92L, 93L, 94L, 95L, 96L, 97L, 98L, 99L, 100L, 
101L, 102L, 103L, 104L, 105L, 106L, 107L, 108L, 109L, 110L, 111L, 
112L, 113L, 114L, 115L, 116L, 117L, 118L, 119L, 120L, 121L, 122L, 
123L, 124L, 125L, 126L, 127L, 128L, 129L, 130L, 131L, 132L, 133L, 
134L, 135L, 136L, 137L, 138L, 139L, 140L, 141L, 142L, 143L, 144L, 
145L, 146L, 147L, 148L, 149L, 150L, 151L, 152L, 153L, 154L, 155L, 
156L, 157L, 158L, 159L, 160L, 161L, 162L, 163L, 164L, 165L, 166L, 
167L, 168L, 169L, 170L, 171L, 172L, 173L, 174L, 175L, 176L, 177L, 
178L, 179L, 180L, 181L, 182L, 183L, 184L, 185L, 186L, 187L, 188L, 
189L, 190L, 191L, 192L, 193L, 194L, 195L, 196L, 197L, 198L, 199L, 
200L, 1L, 2L, 3L, 4L, 5L, 6L, 7L, 8L, 9L, 10L, 11L, 12L, 13L, 
14L, 15L, 16L, 17L, 18L, 19L, 20L, 21L, 22L, 23L, 24L, 25L, 26L, 
27L, 28L, 29L, 30L, 31L, 32L, 33L, 34L, 35L, 36L, 37L, 38L, 39L, 
40L, 41L, 42L, 43L, 44L, 45L, 46L, 47L, 48L, 49L, 50L, 51L, 52L, 
53L, 54L, 55L, 56L, 57L, 58L, 59L, 60L, 61L, 62L, 63L, 64L, 65L, 
66L, 67L, 68L, 69L, 70L, 71L, 72L, 73L, 74L, 75L, 76L, 77L, 78L, 
79L, 80L, 81L, 82L, 83L, 84L, 85L, 86L, 87L, 88L, 89L, 90L, 91L, 
92L, 93L, 94L, 95L, 96L, 97L, 98L, 99L, 100L, 101L, 102L, 103L, 
104L, 105L, 106L, 107L, 108L, 109L, 110L, 111L, 112L, 113L, 114L, 
115L, 116L, 117L, 118L, 119L, 120L, 121L, 122L, 123L, 124L, 125L, 
126L, 127L, 128L, 129L, 130L, 131L, 132L, 133L, 134L, 135L, 136L, 
137L, 138L, 139L, 140L, 141L, 142L, 143L, 144L, 145L, 146L, 147L, 
148L, 149L, 150L, 151L, 152L, 153L, 154L, 155L, 156L, 157L, 158L, 
159L, 160L, 161L, 162L, 163L, 164L, 165L, 166L, 167L, 168L, 169L, 
170L, 171L, 172L, 173L, 174L, 175L, 176L, 177L, 178L, 179L, 180L, 
181L, 182L, 183L, 184L, 185L, 186L, 187L, 188L, 189L, 190L, 191L, 
192L, 193L, 194L, 195L, 196L, 197L, 198L, 199L, 200L), group = c(0L, 
0L, 0L, 0L, 0L, 0L, 0L, 0L, 0L, 0L, 0L, 0L, 0L, 0L, 0L, 0L, 0L, 
0L, 0L, 0L, 0L, 0L, 0L, 0L, 0L, 0L, 0L, 0L, 0L, 0L, 0L, 0L, 0L, 
0L, 0L, 0L, 0L, 0L, 0L, 0L, 0L, 0L, 0L, 0L, 0L, 0L, 0L, 0L, 0L, 
0L, 1L, 1L, 1L, 1L, 1L, 1L, 1L, 1L, 1L, 1L, 1L, 1L, 1L, 1L, 1L, 
1L, 1L, 1L, 1L, 1L, 1L, 1L, 1L, 1L, 1L, 1L, 1L, 1L, 1L, 1L, 1L, 
1L, 1L, 1L, 1L, 1L, 1L, 1L, 1L, 1L, 1L, 1L, 1L, 1L, 1L, 1L, 1L, 
1L, 1L, 1L, 2L, 2L, 2L, 2L, 2L, 2L, 2L, 2L, 2L, 2L, 2L, 2L, 2L, 
2L, 2L, 2L, 2L, 2L, 2L, 2L, 2L, 2L, 2L, 2L, 2L, 2L, 2L, 2L, 2L, 
2L, 2L, 2L, 2L, 2L, 2L, 2L, 2L, 2L, 2L, 2L, 2L, 2L, 2L, 2L, 2L, 
2L, 2L, 2L, 2L, 2L, 3L, 3L, 3L, 3L, 3L, 3L, 3L, 3L, 3L, 3L, 3L, 
3L, 3L, 3L, 3L, 3L, 3L, 3L, 3L, 3L, 3L, 3L, 3L, 3L, 3L, 3L, 3L, 
3L, 3L, 3L, 3L, 3L, 3L, 3L, 3L, 3L, 3L, 3L, 3L, 3L, 3L, 3L, 3L, 
3L, 3L, 3L, 3L, 3L, 3L, 3L, 0L, 0L, 0L, 0L, 0L, 0L, 0L, 0L, 0L, 
0L, 0L, 0L, 0L, 0L, 0L, 0L, 0L, 0L, 0L, 0L, 0L, 0L, 0L, 0L, 0L, 
0L, 0L, 0L, 0L, 0L, 0L, 0L, 0L, 0L, 0L, 0L, 0L, 0L, 0L, 0L, 0L, 
0L, 0L, 0L, 0L, 0L, 0L, 0L, 0L, 0L, 1L, 1L, 1L, 1L, 1L, 1L, 1L, 
1L, 1L, 1L, 1L, 1L, 1L, 1L, 1L, 1L, 1L, 1L, 1L, 1L, 1L, 1L, 1L, 
1L, 1L, 1L, 1L, 1L, 1L, 1L, 1L, 1L, 1L, 1L, 1L, 1L, 1L, 1L, 1L, 
1L, 1L, 1L, 1L, 1L, 1L, 1L, 1L, 1L, 1L, 1L, 2L, 2L, 2L, 2L, 2L, 
2L, 2L, 2L, 2L, 2L, 2L, 2L, 2L, 2L, 2L, 2L, 2L, 2L, 2L, 2L, 2L, 
2L, 2L, 2L, 2L, 2L, 2L, 2L, 2L, 2L, 2L, 2L, 2L, 2L, 2L, 2L, 2L, 
2L, 2L, 2L, 2L, 2L, 2L, 2L, 2L, 2L, 2L, 2L, 2L, 2L, 3L, 3L, 3L, 
3L, 3L, 3L, 3L, 3L, 3L, 3L, 3L, 3L, 3L, 3L, 3L, 3L, 3L, 3L, 3L, 
3L, 3L, 3L, 3L, 3L, 3L, 3L, 3L, 3L, 3L, 3L, 3L, 3L, 3L, 3L, 3L, 
3L, 3L, 3L, 3L, 3L, 3L, 3L, 3L, 3L, 3L, 3L, 3L, 3L, 3L, 3L), 
    tumour = c(0L, 0L, 0L, 0L, 0L, 0L, 0L, 0L, 0L, 0L, 0L, 0L, 
    0L, 0L, 0L, 0L, 0L, 0L, 0L, 0L, 0L, 0L, 0L, 0L, 0L, 0L, 0L, 
    0L, 0L, 0L, 0L, 0L, 0L, 0L, 0L, 0L, 0L, 0L, 0L, 0L, 0L, 0L, 
    0L, 0L, 0L, 0L, 0L, 0L, 0L, 0L, 0L, 0L, 0L, 1L, 1L, 0L, 0L, 
    0L, 0L, 0L, 0L, 0L, 0L, 0L, 0L, 0L, 0L, 0L, 0L, 1L, 0L, 1L, 
    0L, 0L, 0L, 0L, 0L, 0L, 0L, 0L, 0L, 1L, 1L, 0L, 0L, 0L, 0L, 
    0L, 0L, 0L, 0L, 0L, 0L, 0L, 0L, 0L, 0L, 1L, 1L, 1L, 0L, 0L, 
    0L, 0L, 0L, 0L, 0L, 0L, 0L, 0L, 0L, 0L, 0L, 1L, 0L, 0L, 1L, 
    0L, 0L, 0L, 0L, 1L, 0L, 0L, 1L, 0L, 0L, 0L, 0L, 1L, 0L, 0L, 
    0L, 0L, 1L, 0L, 0L, 0L, 0L, 0L, 0L, 0L, 0L, 0L, 0L, 0L, 0L, 
    0L, 1L, 1L, 0L, 0L, 0L, 0L, 0L, 0L, 0L, 0L, 0L, 0L, 0L, 0L, 
    0L, 0L, 0L, 0L, 0L, 1L, 0L, 0L, 0L, 0L, 1L, 0L, 0L, 0L, 0L, 
    0L, 0L, 0L, 0L, 0L, 0L, 0L, 0L, 1L, 1L, 0L, 0L, 0L, 0L, 0L, 
    1L, 0L, 0L, 0L, 0L, 0L, 0L, 0L, 1L, 0L, 0L, 0L, 0L, 0L, 0L, 
    0L, 0L, 0L, 0L, 0L, 0L, 0L, 0L, 0L, 0L, 0L, 0L, 0L, 0L, 0L, 
    0L, 0L, 0L, 0L, 0L, 0L, 0L, 0L, 0L, 0L, 0L, 0L, 0L, 0L, 0L, 
    0L, 0L, 0L, 0L, 0L, 0L, 0L, 0L, 0L, 0L, 0L, 0L, 1L, 0L, 0L, 
    0L, 1L, 1L, 0L, 0L, 0L, 0L, 0L, 0L, 0L, 0L, 0L, 0L, 0L, 0L, 
    0L, 0L, 1L, 0L, 1L, 0L, 0L, 0L, 0L, 0L, 0L, 0L, 0L, 0L, 1L, 
    1L, 0L, 0L, 0L, 0L, 0L, 0L, 0L, 0L, 0L, 0L, 0L, 0L, 0L, 0L, 
    1L, 1L, 1L, 0L, 0L, 0L, 0L, 0L, 0L, 0L, 0L, 0L, 0L, 0L, 0L, 
    0L, 1L, 0L, 0L, 1L, 0L, 0L, 0L, 0L, 1L, 0L, 0L, 1L, 0L, 0L, 
    0L, 0L, 1L, 0L, 0L, 0L, 0L, 1L, 0L, 0L, 0L, 0L, 0L, 0L, 0L, 
    0L, 0L, 0L, 0L, 0L, 0L, 1L, 1L, 0L, 0L, 0L, 0L, 0L, 0L, 0L, 
    0L, 0L, 0L, 0L, 0L, 0L, 0L, 0L, 0L, 0L, 0L, 0L, 0L, 0L, 0L, 
    1L, 0L, 0L, 0L, 0L, 0L, 0L, 0L, 0L, 0L, 0L, 0L, 0L, 1L, 1L, 
    0L, 0L, 0L, 0L, 0L, 1L, 0L, 0L, 0L, 0L, 0L, 0L, 0L), death = c(344L, 
    521L, 529L, 553L, 564L, 588L, 603L, 610L, 610L, 614L, 617L, 
    626L, 639L, 642L, 652L, 661L, 662L, 668L, 704L, 704L, 707L, 
    707L, 712L, 714L, 715L, 718L, 721L, 722L, 722L, 729L, 730L, 
    730L, 730L, 730L, 730L, 730L, 730L, 730L, 730L, 730L, 730L, 
    730L, 730L, 730L, 730L, 730L, 730L, 730L, 730L, 730L, 403L, 
    406L, 431L, 535L, 542L, 572L, 575L, 579L, 579L, 589L, 591L, 
    596L, 602L, 606L, 625L, 638L, 639L, 650L, 660L, 661L, 664L, 
    673L, 674L, 680L, 695L, 703L, 704L, 704L, 704L, 712L, 718L, 
    718L, 718L, 725L, 730L, 730L, 730L, 730L, 730L, 730L, 730L, 
    730L, 730L, 730L, 730L, 730L, 730L, 730L, 730L, 730L, 15L, 
    438L, 467L, 502L, 521L, 522L, 568L, 572L, 582L, 595L, 596L, 
    601L, 610L, 619L, 630L, 638L, 639L, 642L, 642L, 651L, 658L, 
    659L, 661L, 664L, 673L, 674L, 692L, 695L, 695L, 695L, 696L, 
    696L, 706L, 709L, 712L, 730L, 730L, 730L, 730L, 730L, 730L, 
    730L, 730L, 730L, 730L, 730L, 730L, 730L, 730L, 730L, 337L, 
    409L, 457L, 467L, 495L, 502L, 523L, 546L, 547L, 568L, 575L, 
    583L, 584L, 598L, 598L, 600L, 602L, 607L, 610L, 614L, 621L, 
    625L, 633L, 638L, 642L, 642L, 642L, 646L, 648L, 650L, 654L, 
    658L, 659L, 660L, 660L, 660L, 660L, 661L, 669L, 670L, 680L, 
    680L, 683L, 684L, 688L, 688L, 699L, 700L, 704L, 712L, 144L, 
    521L, 529L, 553L, 564L, 588L, 603L, 610L, 610L, 614L, 617L, 
    626L, 639L, 642L, 652L, 661L, 662L, 668L, 704L, 704L, 707L, 
    707L, 712L, 714L, 715L, 718L, 721L, 722L, 722L, 729L, 730L, 
    730L, 730L, 730L, 730L, 730L, 730L, 730L, 730L, 730L, 730L, 
    730L, 730L, 730L, 730L, 730L, 730L, 730L, 730L, 730L, 203L, 
    406L, 431L, 535L, 542L, 572L, 575L, 579L, 579L, 589L, 591L, 
    596L, 602L, 606L, 625L, 638L, 639L, 650L, 660L, 661L, 664L, 
    673L, 674L, 680L, 695L, 703L, 704L, 704L, 704L, 712L, 718L, 
    718L, 718L, 725L, 730L, 730L, 730L, 730L, 730L, 730L, 730L, 
    730L, 730L, 730L, 730L, 730L, 730L, 730L, 730L, 730L, 15L, 
    438L, 467L, 502L, 521L, 522L, 568L, 572L, 582L, 595L, 596L, 
    601L, 610L, 619L, 630L, 638L, 639L, 642L, 642L, 651L, 658L, 
    659L, 661L, 664L, 673L, 674L, 692L, 695L, 695L, 695L, 696L, 
    696L, 706L, 709L, 712L, 730L, 730L, 730L, 730L, 730L, 730L, 
    730L, 730L, 730L, 730L, 730L, 730L, 730L, 730L, 730L, 337L, 
    409L, 457L, 467L, 495L, 502L, 523L, 546L, 547L, 568L, 575L, 
    583L, 584L, 598L, 598L, 600L, 602L, 607L, 610L, 614L, 621L, 
    625L, 403L, 638L, 642L, 642L, 642L, 646L, 648L, 650L, 654L, 
    658L, 659L, 660L, 660L, 660L, 660L, 661L, 669L, 670L, 680L, 
    680L, 683L, 684L, 688L, 688L, 699L, 700L, 704L, 712L), study = structure(c(1L, 
    1L, 1L, 1L, 1L, 1L, 1L, 1L, 1L, 1L, 1L, 1L, 1L, 1L, 1L, 1L, 
    1L, 1L, 1L, 1L, 1L, 1L, 1L, 1L, 1L, 1L, 1L, 1L, 1L, 1L, 1L, 
    1L, 1L, 1L, 1L, 1L, 1L, 1L, 1L, 1L, 1L, 1L, 1L, 1L, 1L, 1L, 
    1L, 1L, 1L, 1L, 1L, 1L, 1L, 1L, 1L, 1L, 1L, 1L, 1L, 1L, 1L, 
    1L, 1L, 1L, 1L, 1L, 1L, 1L, 1L, 1L, 1L, 1L, 1L, 1L, 1L, 1L, 
    1L, 1L, 1L, 1L, 1L, 1L, 1L, 1L, 1L, 1L, 1L, 1L, 1L, 1L, 1L, 
    1L, 1L, 1L, 1L, 1L, 1L, 1L, 1L, 1L, 1L, 1L, 1L, 1L, 1L, 1L, 
    1L, 1L, 1L, 1L, 1L, 1L, 1L, 1L, 1L, 1L, 1L, 1L, 1L, 1L, 1L, 
    1L, 1L, 1L, 1L, 1L, 1L, 1L, 1L, 1L, 1L, 1L, 1L, 1L, 1L, 1L, 
    1L, 1L, 1L, 1L, 1L, 1L, 1L, 1L, 1L, 1L, 1L, 1L, 1L, 1L, 1L, 
    1L, 1L, 1L, 1L, 1L, 1L, 1L, 1L, 1L, 1L, 1L, 1L, 1L, 1L, 1L, 
    1L, 1L, 1L, 1L, 1L, 1L, 1L, 1L, 1L, 1L, 1L, 1L, 1L, 1L, 1L, 
    1L, 1L, 1L, 1L, 1L, 1L, 1L, 1L, 1L, 1L, 1L, 1L, 1L, 1L, 1L, 
    1L, 1L, 1L, 1L, 2L, 2L, 2L, 2L, 2L, 2L, 2L, 2L, 2L, 2L, 2L, 
    2L, 2L, 2L, 2L, 2L, 2L, 2L, 2L, 2L, 2L, 2L, 2L, 2L, 2L, 2L, 
    2L, 2L, 2L, 2L, 2L, 2L, 2L, 2L, 2L, 2L, 2L, 2L, 2L, 2L, 2L, 
    2L, 2L, 2L, 2L, 2L, 2L, 2L, 2L, 2L, 2L, 2L, 2L, 2L, 2L, 2L, 
    2L, 2L, 2L, 2L, 2L, 2L, 2L, 2L, 2L, 2L, 2L, 2L, 2L, 2L, 2L, 
    2L, 2L, 2L, 2L, 2L, 2L, 2L, 2L, 2L, 2L, 2L, 2L, 2L, 2L, 2L, 
    2L, 2L, 2L, 2L, 2L, 2L, 2L, 2L, 2L, 2L, 2L, 2L, 2L, 2L, 2L, 
    2L, 2L, 2L, 2L, 2L, 2L, 2L, 2L, 2L, 2L, 2L, 2L, 2L, 2L, 2L, 
    2L, 2L, 2L, 2L, 2L, 2L, 2L, 2L, 2L, 2L, 2L, 2L, 2L, 2L, 2L, 
    2L, 2L, 2L, 2L, 2L, 2L, 2L, 2L, 2L, 2L, 2L, 2L, 2L, 2L, 2L, 
    2L, 2L, 2L, 2L, 2L, 2L, 2L, 2L, 2L, 2L, 2L, 2L, 2L, 2L, 2L, 
    2L, 2L, 2L, 2L, 2L, 2L, 2L, 2L, 2L, 2L, 2L, 2L, 2L, 2L, 2L, 
    2L, 2L, 2L, 2L, 2L, 2L, 2L, 2L, 2L, 2L, 2L, 2L, 2L, 2L, 2L, 
    2L, 2L, 2L, 2L, 2L, 2L, 2L, 2L, 2L), .Label = c("a", "b"), class = "factor"), 
    dose = structure(c(1L, 1L, 1L, 1L, 1L, 1L, 1L, 1L, 1L, 1L, 
    1L, 1L, 1L, 1L, 1L, 1L, 1L, 1L, 1L, 1L, 1L, 1L, 1L, 1L, 1L, 
    1L, 1L, 1L, 1L, 1L, 1L, 1L, 1L, 1L, 1L, 1L, 1L, 1L, 1L, 1L, 
    1L, 1L, 1L, 1L, 1L, 1L, 1L, 1L, 1L, 1L, 2L, 2L, 2L, 2L, 2L, 
    2L, 2L, 2L, 2L, 2L, 2L, 2L, 2L, 2L, 2L, 2L, 2L, 2L, 2L, 2L, 
    2L, 2L, 2L, 2L, 2L, 2L, 2L, 2L, 2L, 2L, 2L, 2L, 2L, 2L, 2L, 
    2L, 2L, 2L, 2L, 2L, 2L, 2L, 2L, 2L, 2L, 2L, 2L, 2L, 2L, 2L, 
    3L, 3L, 3L, 3L, 3L, 3L, 3L, 3L, 3L, 3L, 3L, 3L, 3L, 3L, 3L, 
    3L, 3L, 3L, 3L, 3L, 3L, 3L, 3L, 3L, 3L, 3L, 3L, 3L, 3L, 3L, 
    3L, 3L, 3L, 3L, 3L, 3L, 3L, 3L, 3L, 3L, 3L, 3L, 3L, 3L, 3L, 
    3L, 3L, 3L, 3L, 3L, 4L, 4L, 4L, 4L, 4L, 4L, 4L, 4L, 4L, 4L, 
    4L, 4L, 4L, 4L, 4L, 4L, 4L, 4L, 4L, 4L, 4L, 4L, 4L, 4L, 4L, 
    4L, 4L, 4L, 4L, 4L, 4L, 4L, 4L, 4L, 4L, 4L, 4L, 4L, 4L, 4L, 
    4L, 4L, 4L, 4L, 4L, 4L, 4L, 4L, 4L, 4L, 1L, 1L, 1L, 1L, 1L, 
    1L, 1L, 1L, 1L, 1L, 1L, 1L, 1L, 1L, 1L, 1L, 1L, 1L, 1L, 1L, 
    1L, 1L, 1L, 1L, 1L, 1L, 1L, 1L, 1L, 1L, 1L, 1L, 1L, 1L, 1L, 
    1L, 1L, 1L, 1L, 1L, 1L, 1L, 1L, 1L, 1L, 1L, 1L, 1L, 1L, 1L, 
    2L, 2L, 2L, 2L, 2L, 2L, 2L, 2L, 2L, 2L, 2L, 2L, 2L, 2L, 2L, 
    2L, 2L, 2L, 2L, 2L, 2L, 2L, 2L, 2L, 2L, 2L, 2L, 2L, 2L, 2L, 
    2L, 2L, 2L, 2L, 2L, 2L, 2L, 2L, 2L, 2L, 2L, 2L, 2L, 2L, 2L, 
    2L, 2L, 2L, 2L, 2L, 3L, 3L, 3L, 3L, 3L, 3L, 3L, 3L, 3L, 3L, 
    3L, 3L, 3L, 3L, 3L, 3L, 3L, 3L, 3L, 3L, 3L, 3L, 3L, 3L, 3L, 
    3L, 3L, 3L, 3L, 3L, 3L, 3L, 3L, 3L, 3L, 3L, 3L, 3L, 3L, 3L, 
    3L, 3L, 3L, 3L, 3L, 3L, 3L, 3L, 3L, 3L, 4L, 4L, 4L, 4L, 4L, 
    4L, 4L, 4L, 4L, 4L, 4L, 4L, 4L, 4L, 4L, 4L, 4L, 4L, 4L, 4L, 
    4L, 4L, 4L, 4L, 4L, 4L, 4L, 4L, 4L, 4L, 4L, 4L, 4L, 4L, 4L, 
    4L, 4L, 4L, 4L, 4L, 4L, 4L, 4L, 4L, 4L, 4L, 4L, 4L, 4L, 4L
    ), .Label = c("0", "1", "2", "3"), class = "factor")), .Names = c("no", 
"group", "tumour", "death", "study", "dose"), row.names = c(NA, 
-400L), class = "data.frame")
@


<<polY1, echo=TRUE, results='asis', message=FALSE, warning=FALSE>>=
library(lme4)
library(multcomp)
meLH$dose<-as.factor(meLH$group) # factor
# study-specific poly3 weights
meLHa<- meLH[meLH$study=="a", ]
meLHa$weightpoly3 <- 1 # Compute the poly-3 (-k)- weights at the level of single animals
wt0 <- which(meLHa$tumour == 0)
meLHa$weightpoly3[wt0] <- (meLHa$death[wt0]/max(meLHa$death))^3
meLHb<- meLH[meLH$study=="b", ]
meLHb$weightpoly3 <- 1 
wt0 <- which(meLHb$tumour == 0)
meLHb$weightpoly3[wt0] <- (meLHb$death[wt0]/max(meLHb$death))^3
ME<-rbind(meLHa,meLHb)
# glmm with weights
mo2 <- glmer(cbind(tumour,50-tumour) ~ dose-1 +(1|study),family = binomial,
             weight=weightpoly3, data=ME)
m2<-glht(mo2, linfct = mcp(dose = "Dunnett"), alternative = "greater" )
SM2<-summary(m2)

@

<<repP, echo=FALSE,results='asis', warning=FALSE, message=FALSE>>=
library("xtable")
library("ggplot2")
Tboli<-fortify(summary(SM2))[, c(1,5,6)]
colnames(Tboli)<-c("Model","Test stats", "p-value")
print(xtable(Tboli, digits=c(1,1,2,7), caption="Dunnett-type test for poly3 estimates using a mixed effect model", 
             label="tab:boli"), include.rownames=FALSE)
@
\normalsize
\end{verbatim}

\end{document}